# Selection of Proper EEG Channels for Subject Intention Classification Using Deep Learning

Ghazale Ghorbanzade[1], Zahra Nabizadeh-ShahreBabak[1], Shadrokh Samavi[1,2], Nader Karimi[1], Ali Emami[1], Pejman Khadivi[3]

[1]Department of Elect and Computer Eng., Isfahan University of Technology, Isfahan, Iran
[2]Department of Elect and Computer Eng., McMaster University, Hamilton, Canada
[3]Computer Science Department, Seattle University, Seattle, USA
khadivip@seattleu.edu

**Abstract.** Brain signals could be used to control devices to assist individuals with disabilities. Signals such as electroencephalograms are complicated and hard to interpret. A set of signals are collected and should be classified to identify the intention of the subject. Different approaches have tried to reduce the number of channels before sending them to a classifier. We are proposing a deep learning-based method for selecting an informative subset of channels that produce high classification accuracy. The proposed network could be trained for an individual subject to select an appropriate set of channels. The channel section could reduce the complexity of brain-computer-interface devices. Our method could find a subset of channels. The accuracy of our approach is comparable with a model trained on all channels. Hence, our model's temporal and power costs are low, while its accuracy is kept high.

## 1 Introduction

Brain-Computer Interface (BCI) is a system that provides the communication of brain signals to peripheral devices such as computers. In these systems, users' intentions could be read by decoding the features of brain signals. The signals could be translated into specific commands that control devices such as computers and wheelchairs [1]. Measuring brain activities is the centerpiece in a BCI system. There are different techniques for recording brain activities such as electroencephalogram (EEG), functional magnetic resonance imaging (fMRI), and functional near-infrared spectroscopy (fNIRS). Among these techniques, EEG and fNIRS methods use portable equipment, which makes them suitable for BCI systems. In recent research works, the following types of EEG signals are used:

- Motor Imagery (MI): this is the most common research on EEG signals where they are recorded while imagining a specific movement. For example, the subject imagines that her right-hand moves without performing any physical activity.

- Steady-State Visual Evoked Potentials (SSVEP): this is a type of EEG signal in response to visual stimulus.
- Event-Related Potential (ERP): in this type, the brain's response is recorded during a task-relevant stimulus. The most important signal in this type is P300, one of the most commonly detected and reliable signals.

The signals recorded with these methods could be analyzed with signal processing, machine learning, and neural networks methods.

One main task for translating brain signals to commands is classification. Recently, deep learning (DL) methods have attracted lots of attention. In [2], a one-dimensional convolutional neural network (CNN) is combined with a long short term memory (LSTM) to classify motor imagery (MI) tasks. Due to EEG signals' time-series nature, using LSTM structures would extract better features and achieve better accuracy [2].

In [3], J. Lawhern et al. proposed a CNN architecture that could generally be used for most BCI paradigms. They believe that rather than using general-purpose convolutional networks, it is best to use custom made structures, and that is due to EEG signals' specific characteristics. For this purpose, they proposed a compact network called EEGNet. In this architecture, they used depth-wise and separable convolutions to extract EEG features for BCI applications. They evaluated their network in four BCI paradigms, P300 visual-evoked potentials, error-related negativity responses (ERN), movement-related cortical potentials (MRCP), and sensory-motor rhythms (SMR). For these situations, they demonstrated promising results [3]. For achieving high classification accuracy, one has to increase the size of the models by increasing the number of layers and filters. In this situation, extensive memory and computational resources are required. Extensive use of resources is not desirable because the main challenge in an embedded BCI system is low power consumption.

Ingolfsson et al., in [4], proposed a new network architecture with few trainable parameters. This network was named EEG-TCNET, a temporal convolutional network (TCN), with low memory and low computational complexity, making it suitable for embedded classification. In [5], Amin et al. believe that due to the low signal to noise ratio of the EEG signal and its dynamic nature, its decoding is complicated and hard. Hence, they suggest DL models for this work. For this purpose, multiple CNN models are used for extracting features, which are fused and applied for classification.

In some research works, signal processing methods are used for feature extraction. Machine learning methods, such as support vector machine (SVM), are used for classification. In [6], two methods, common spatial pattern (CSP) and Riemannian covariance methods, are selected to extract features. EEG signals extend these two methods to make them suitable for multiscale temporal and spectral cases and apply SVM for classification. By using these methods, the execution time during training and test is decreased.

The previously mentioned research works, introducing new network architectures, feature extraction methods, and classification approaches showing improvements in accuracy, complexity, and speed.

Another research trend focuses on reducing the number of input channels by selecting more informative ones to increase accuracy and speed and reduce its complexity. In [7], the binary gravitation search algorithm (BGSA) is utilized for selecting a subset of optimal channels. In their work, a bandpass filter is applied to reduce the noise. Also, they use the blind source separation (BSS) algorithm to remove artifacts. Artifacts include electrooculography (EOG) and electromyography (EMG). Then for extracting features, signals are analyzed in time and wavelet domain. Finally, the improved BGSA (IBGSA) is used for searching optimal channels. The efficiency of their channel selection method is evaluated by SVM.

In [8], correlation coefficient values are used for selecting distinctive channels. The distinctive channels are selected by estimating uncorrelated channels based on correlation coefficients. For each channel, a group of uncorrelated channels is selected. For choosing one of these groups, the filter-bank CSP (FBCSP) is initially applied to each group for extracting features, and then the Fisher score is calculated. The group with the highest Fisher score is selected. Qi et al. [9] proposed a channel selection method using spatiotemporal information of EEG data. They defined channel selection as an optimization problem and developed a computationally efficient algorithm for solving it by incorporating sparsity constraints [9]. Using CSP is one of the methods for channel selection. However, it is highly sensitive to the frequency band and time window of EEG segments. In [10], Zhang et al. introduce a new method for improving CSP, named temporally constrained sparse group spatial pattern (TSGSP). In this method, the effect of frequency bands and time windows are considered simultaneously.

In this paper, we propose a DL-based channel selection method, which could select an informative set of channels for classification applications. Due to the complexity of EEG datasets, deep learning models could result in a better selection of channels compared to machine learning or signal processing methods.

The rest of the paper is organized as follows. In section II, the proposed method is explained. Experimental results are reported in section III, and section IV represents concluding remarks.

## 2 Proposed Method

Recent EEG datasets contain more channels to provide higher spatial resolution systems to achieve more accurate models. However, dealing with more channels does not necessarily ensure better classification. This leads to an increase in the network input size, making model training a more complicated process. It also demands more hardware resources, processing time, and larger datasets to train a proper model. Channel selection is a common approach [4, 7, 8] in which the input dimensions are reduced, and suitable channels for the task are chosen.

In traditional single-trial EEG classification methods, the channels are selected based on appropriate criteria for feature extraction. However, in the classification of EEG signals based on deep learning, determining channel selection criteria is not simple. This is because the features in these methods are extracted using the learning process. Therefore, the nature of the extracted features is not identifiable, and therefore selecting an appropriate criterion is not easy.

In this paper, we discuss several learning-based channel selection methods to view the problem from different perspectives. For this purpose, we consider various approaches for selecting candidate subsets, which are fed into the desired classification network. The trained models' accuracy on each subset of channels is compared with each other, and the best subset is selected. We suggest channel selection based on experimental results, which reduces the computational cost and leads to high accuracy comparable to using all EEG channels.

Let us assume a dataset containing $N$ samples denoted by $\mathcal{R} = \{R_1, R_2, \dots, R_N\}$. Every sample $R_i$ in the dataset belongs to $\mathbb{R}^{C \times T}$, where $'C'$ is channel numbers and $'T'$ is the number of time samples in EEG signal. Also, assume channel set is determined as $S = \{s_1, s_2, \dots, s_C\}$ where $s_i$ represents channel $i$. For channel subset $S'$ with dimension $C'$, a modified dataset is created by choosing channels that exist in $S'$. This dataset is denoted as $\mathcal{R}'$ and its data samples, $R_i'$ belongs to $\mathbb{R}^{C' \times T}$. For each subset, the relevant dataset is built and used to train the model. The model's accuracy is used to compare selected subsets and determine how effective the channel selection process has been. Here, the network used to compare the accuracy of channel subsets is EEGNet.

An intuitive approach to finding a subset of channels that gives the best accuracy on the training data is an exhaustive search. One can obtain all possible subsets of channels and select the subset with the highest accuracy. This full search method is not economical in terms of time and consumption of resources. Therefore, it makes more sense to use a more efficient and sub-optimal method instead. For this purpose, two sub-optimal methods are used to search the subset space of channels. Another view is selecting channels based on the task in which signals are recorded. In this paper, the MI task is considered, so selecting channels may be useful based on the task.

In short, we investigate and compare three different methods for the selection of proper EEG subset 1. Incremental search, 2. Weighted random search, and 3. Task-based channel selection. The following is the description of each of these methods.

### 2.1 Hierarchical incremental search

The main idea in the incremental search approach is to start with the most informative channel and add one most significant channel at each stage. In this algorithm, every channel is fed to the network, and the training process is performed. Then the channel, which leads to the best accuracy, is selected. In the next step, the chosen channel is combined with other channels to create two-channel subsets. For each subset, a model is trained and evaluated. Similarly, among these two-channel subsets, the subset corresponding with the most accurate model is selected. This procedure is performed similarly in every step, and it continues until the complete set of channels is used. Finally, for any number of channels, the accuracy of the best subset is reported. The subset with the most accurate model among these subsets is specified as the selected channel.

This algorithm is a sub-optimal method, and there may be better subsets that are not considered.

### 2.2 Weighted random search

The hierarchical structure of the mentioned methods leads to inflexibility in the channel selection method, and one hierarchy may be selected. Simultaneously, another one with a little less accuracy in the middle makes it more accurate if it goes to the next steps.

Adding random nature to channel selection could lead to higher performance because the selection is not limited to a specific hierarchy. To this end, firstly, the proposed method selects $k$ random subset with equal probability for selecting or not selecting each channel. For every subset, a model is trained, and its accuracy is kept as the weights of existing channels in the corresponding subset. For each subset, a one-hot vector $\underline{P_j}, j \in \{1, \dots, k\}$ is determined so that

$$\underline{P_j} = \begin{bmatrix} p_{j1} \\ \vdots \\ p_{jC} \end{bmatrix} , p_{ji} = \begin{cases} 1 \;\; if\; i^{th}\; channel\; exist\; in\; j^{th} subset \\ 0 \;\; if\; i^{th}\; channel\; doesn't\; exist\; in\; j^{th}\; subset \end{cases} \quad (1)$$

By denoting accuracy of each subset as $w_j$, scores of each channel is calculated based on Eq. 2.

$$v_i = \sum_{j=1}^{k} p_{ji} w_j \quad (2)$$

Finally, channels with higher scores would be more informative. So the desired channels are selected based on obtained scores.
If the number of selecting random subsets is large enough, a reasonable estimate of each channel score will be obtained.

### 2.3 Task-based channel selection

Although the brain structure is very complex, studies show that each part of the brain has more responsibility for specific tasks [11]. This does not mean that only that part of the brain is responsible for that task or that the only activity performed by this part of the brain is that task. Instead, while performing this task, the brain's function has changed more in its assigned part. According to this, the subset of channels, which are located on the motor cortex, the part of the brain responsible for motor imagery tasks, based on 10-20 standard of EEG channels is selected, and the results would be compared with other methods.

## 3 Experimental Results

As mentioned before, in this work, the goal was to propose a deep learning-based channel selection approach that decreased time and resource costs in the test phase and increased accuracy or at least without much drop in performance. At first, the dataset used to evaluate implementations, BCI Competition dataset IV 2a, is described. Then, the implementation details and results of each of the mentioned methods are explained. In the end, the proposed method's performance is compared with others in terms of accuracy and number of model parameters.

### 3.1 Implementation Details

The implementation process is done on a system with a Core i7 CPU, 32GB RAM, and a GeForce RTX 2080 Ti GPU. Also, the software that we used is Keras API, with the backend of TensorFlow. EEG-Net is utilized as the training process model, and it uses accuracy as the evaluation metric and categorical cross-entropy as the loss function. The first-day data of the BCI Competition IV 2a dataset is used as training data, and the data of the second day is used as a test set.

### 3.2 Dataset Description

The dataset used in this work is the BCI-Competition IV 2a dataset, which is publicly available[1]. It consists of nine subject's EEG signals during a four-class motor imagery task. For each subject, the data is recorded during two different days, and in the data of each day, 288 trials exist. Sampling frequency in recording EEG signals is 250 Hz, and a bandpass filter between 0.5 and 100 Hz and a 50 Hz notch filter is applied to it. The duration used in this work starts at 0.5 seconds before the MI cue and continues until the end of the MI, resulting in 4.5 seconds, including 1125 samples of time. The data includes 22 EEG channels, montaged corresponding to the international 10-20 systems. Motor imagery tasks are the imagination of movement of the left hand (class 1), right hand (class 2), both feet (class 3), and tongue (class 4). Trials containing artifacts are removed based on the artifact label in the dataset and marked by experts. In the following, the data of two subjects, subjects 1 and 3, are used to evaluate implemented methods.

### 3.3 Task-Based Channel Selection

As the task defined in the dataset is motor imagery, the electrodes on the motor cortex seem to be informative, and so these electrodes are selected. The selected channels are shown in Fig.1. For this selected subset, the results are presented in Table 1.
Table 1 shows, other methods result in better accuracy compare to this method. It shows that this brain task is not just processed in a specific part, but different parts of the brain are involved in processing this task.

### 3.4 Hierarchical Approaches

Fig. 2 shows a hierarchical aggregation approach for subject 3. It has already been mentioned that in this approach, the followed procedure is started by finding the most informative channel, and in each step, the channel that adding it with the previous subset leads to the best result is selected.

---

[1] http://www.bbci.de/competition/iv/

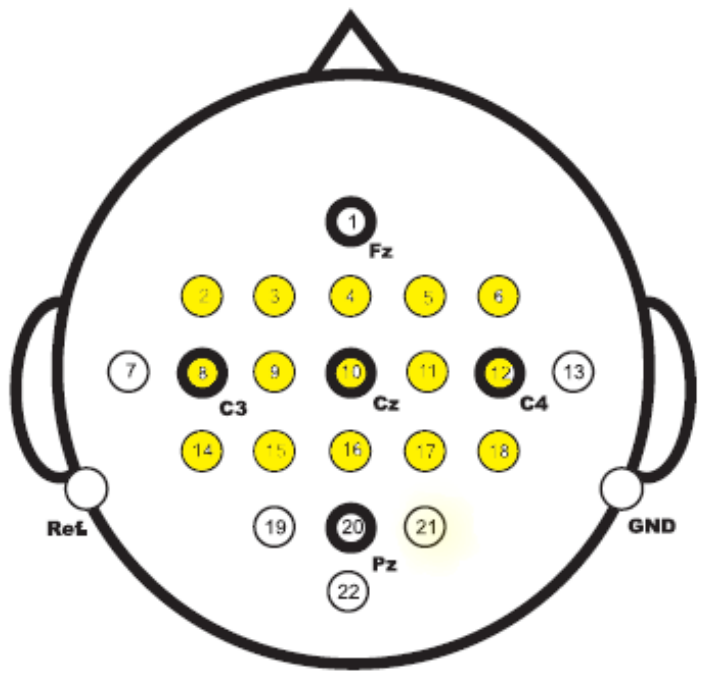

**Fig. 1.** Locations of electrodes on the motor cortex are highlighted in yellow. Electrode montage is according to the 10-20 standard.

**Table 1**. Results of the proposed methods for subsets consisting of 15 channels.

| | **Motor imagery subset** | | **Hierarchical increasing** | | **Method based on channel histogram** | |
|---|---|---|---|---|---|---|
| | **Accuracy** | **Loss** | **Accuracy** | **Loss** | **Accuracy** | **Loss** |
| **Subject 1** | 79.36 | 0.51 | **81.99** | **0.49** | 80.07 | 0.49 |
| **Subject 3** | 84.84 | 0.44 | **87.25** | **0.34** | 85.57 | 0.40 |

In Fig. 2, it can be seen that $12^{th}$ channel is selected in the first step, and it means that this channel has the information for the classification of this subject's EEG signals. Vice versa, the last channel, which is added to the selected set of channels, is the $10^{th}$ channel. Therefore, the $10^{th}$ channel adds less information for the desired task and is not very informative.

However, the last steps are not shown in the figure due to space constraints. In Fig. 2, the results of the selected subset of channels are given. In each step, some other subsets are also presented to show that each step's selected channel causes the best performance. As this method searches among many subsets in a sub-optimal manner, its results are superior to other proposed methods.

### 3.5 Method based on weighted channel histogram

In the implementation of this approach, ten random channel subsets are trained on subjects 1 and 3. The accuracy of each subset is then obtained. The selected subsets are visualized in Table 2. Then in the manner explained in Sec. 3, each channel score is calculated and is presented. Finally, channel selection is proposed based on sorted scores.

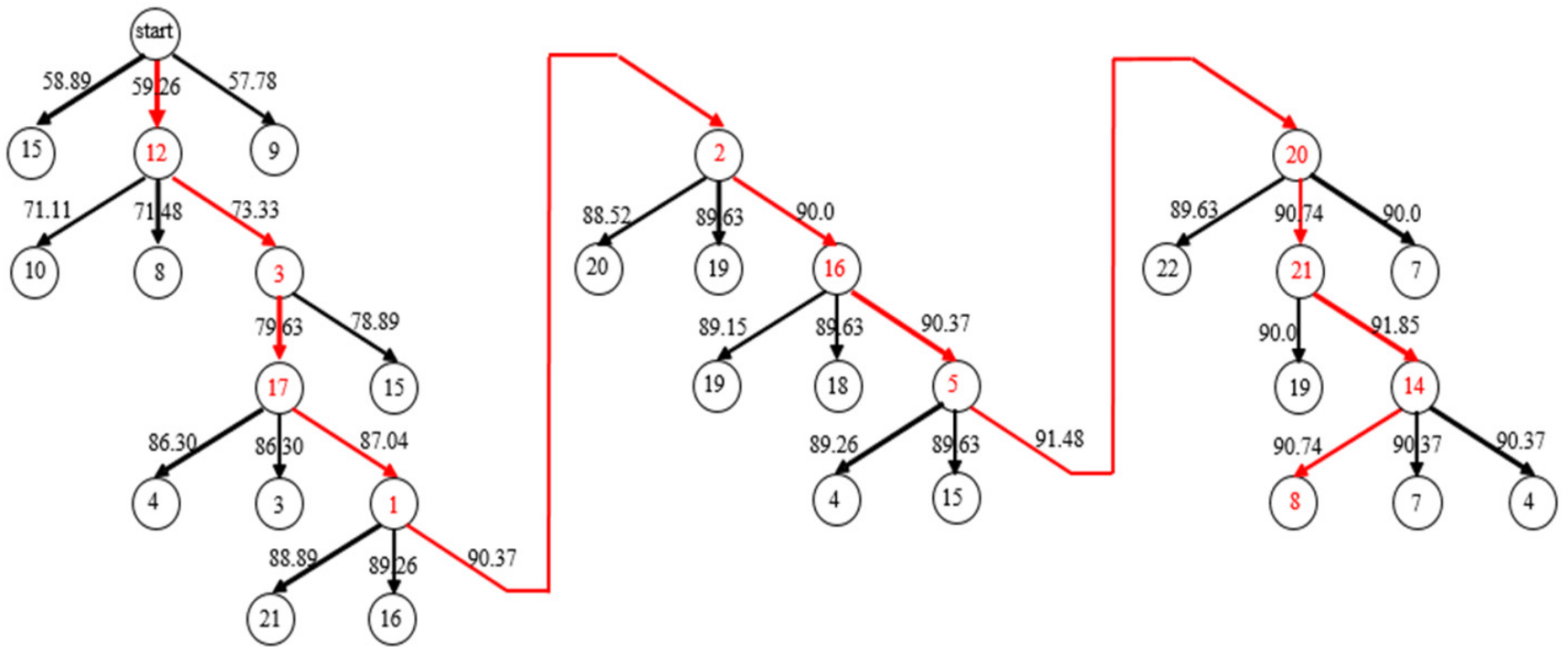

**Fig. 2.** This figure shows how adding a channel is done in each step. Each node shows the channel index, and the number written on its edge represents the model's accuracy after adding that channel to the previous subset.

In the Table. 1, the results of the proposed methods could be seen. Comparing this method with a hierarchical increasing method, the hierarchical method's accuracy in the channel selection procedure is higher. The channel selection is an offline process, and the time it takes does not matter against accuracy. In the current method, the search space is vast for achieving better results than the hierarchical method. Due to this, many training procedures should be done, that it does not make sense.

**Table 2**. The visualization of selected channels in random subsets. The reported accuracy is used as the weight of each channel to calculate its score.

| | | | | |
|---|---|---|---|---|
| 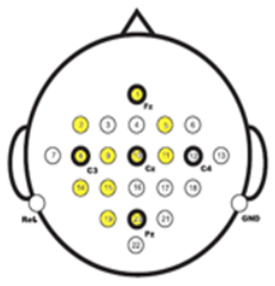 sub1: 78.18% sub3: 83.33% | 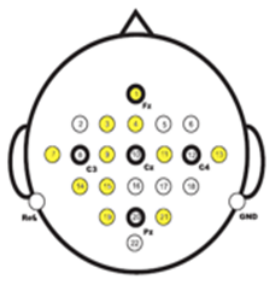 sub1: 70.91% sub3: 82.22% | 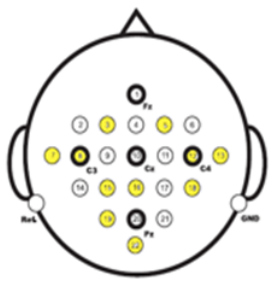 sub1: 78.18% sub3: 86.67% | 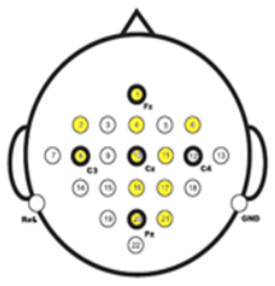 sub1: 77.09% sub3: 86.30% | 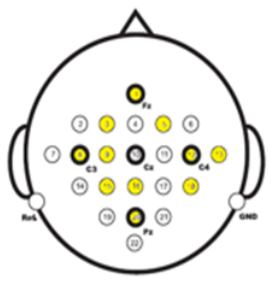 sub1: 83.27% sub3: 87.78% |
| 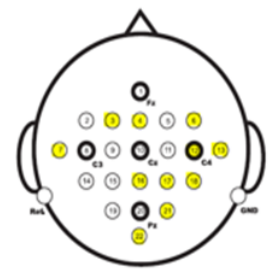 sub1: 73.45% sub3: 84.44% | 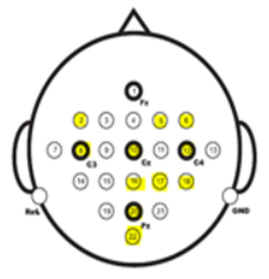 sub1: 83.64% sub3: 85.19% | 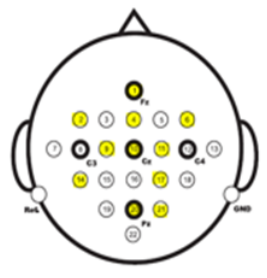 sub1: 72.72% sub3: 80.00% | 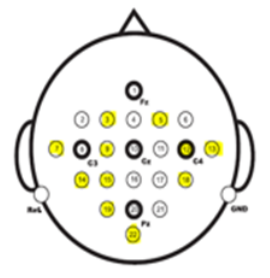 sub1: 77.45 % sub3: 79.26% | 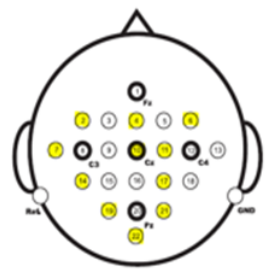 sub1: 70.18% sub3: 81.48% |

### 3.6 Comparing with others

In practical BCI systems, time and resource consumption is one of the significant challenges. On the contrary, the state-of-the-art models use more resources for higher accuracies. In this paper, deep learning-based channel selection methods are used to save resources and time while the accuracy is kept as high as possible. In Table 3, channel selection results are compared with the accuracy reported by [4].

It is seen that using channel selection for EEGNet, improves accuracy. In [4], the number of parameters reported for EEGNet implementation is 2.63 k that is significantly lower than EEG-TCNet with 4.27 k parameters. In our proposed method, using channel selection decreases the input dimension, and as a result, the number of parameters is decreased to 2.50k. Therefore the proposed channel selection reduces the parameters nearly in half ratio compared with EEG-TCnet, while its accuracy is slightly lower. Also, the selected channels by EEGNet, in comparison with EEGNet learned with all channels, have more accuracy for subject 3. The selected channels' performance is slightly lower for subject 1, while its number of parameters is decreased.

**Table 3**. Comparison of the proposed method with others.

| subject | Hierarchical increasing | EEGNet [4] | EEG-TCNet [4] |
|---|---|---|---|
| **Sub1 ($C' = 14$)** | 81.71 | 84.34 | 85.77 |
| **Sub3 ($C' = 14$)** | 89.23 | 87.54 | 94.51 |
| **Parameters** | 2.50k | 2.63k | 4.27k |

## 4 Conclusion

This paper uses a new approach to channel selection, which uses deep learning in the channel selection process. Three methods for channel selection based on deep learning are presented. In the first method, channels assigned to the motor imagery task are chosen. In the second one, a hierarchical incremental search is used to find a proper subset of channels. In the last one, a method based on a weighted channel histogram is utilized to score channels, and the selection of channels is made based on these scores. Among these methods, the hierarchical incremental search shows superiority. This method is a sub-optimal approach to a full search and compares many different subsets to choose a proper one. Compared with applying all channels, using the proposed channel selection method has better accuracy and fewer parameters. Compared with EEG-TCNet, the proposed method uses nearly half its parameters, while accuracy is slightly less.